\documentclass[sigconf]{acmart}
\AtBeginDocument{%
  }

\setcopyright{acmlicensed}
\copyrightyear{2018}
\acmYear{2018}
\acmDOI{XXXXXXX.XXXXXXX}
\acmConference[Conference acronym 'XX]{Make sure to enter the correct
  conference title from your rights confirmation email}{June 03--05,
  2018}{Woodstock, NY}

\acmISBN{978-1-4503-XXXX-X/2018/06}

\usepackage{booktabs}
\usepackage{multirow}
\usepackage{tabularx}
\usepackage{dsfont}

\usepackage{balance}
\usepackage{cleveref}
\usepackage{hyperref}
\hypersetup{colorlinks=true,citecolor=brown}
\usepackage{url}
\usepackage[table,xcdraw,dvipsnames]{xcolor}

\usepackage[most]{tcolorbox}
\usepackage{soul} % line-break-friendly highlighting
\sethlcolor{blue!15}

\newcommand{\inlinehl}[1]{\begingroup\sethlcolor{blue!15}\hl{#1}\endgroup}

\definecolor{refinegreen}{RGB}{0, 128, 75}
\definecolor{scoregreen}{RGB}{34, 139, 34}
\definecolor{scorered}{RGB}{178, 34, 34}

\definecolor{darkgray}{RGB}{70,70,70}
\definecolor{lightgray}{RGB}{240,240,240}

\newcommand{\red}[1]{\textcolor{BrickRed}{#1}}

\usepackage{caption} % for \captionof
\usepackage{enumitem}
\usepackage{needspace}
\newtcolorbox[auto counter, number within=section]{InstructionBox}[2][]{
  enhanced, %breakable,
  colback=gray!3,          % body background
  colframe=gray!70!black,  % frame color
  boxrule=0.5pt,
  arc=2mm,                 % rounded corners
  left=2mm,right=2mm,top=2mm,bottom=2mm,
  drop fuzzy shadow,
  fonttitle=\bfseries\large,
  coltitle=white,
  colbacktitle=black!92,
  attach boxed title to top left={xshift=2mm,yshift=-2mm},
  boxed title style={
    size=small,
    frame hidden,
    interior style={fill=black!92},
    top=1mm,bottom=1mm,left=2mm,right=2mm,
    sharp corners,
  },
title={Instruction \thetcbcounter: #2}, % Automatic numbering + Title
list entry={Instruction \thetcbcounter: #2}, % For the List of Instructions  
#1
}

\newtcolorbox[auto counter, number within=section]{resultbox}[2][]{
  enhanced, %breakable,
  colback=gray!3,
  colbacklower=gray!3,
  segmentation style={draw=none},
  colframe=gray!70!black,  % frame color
  boxrule=0.5pt,
  arc=2mm,                 % rounded corners
  left=2mm,right=2mm,top=2mm,bottom=2mm,
  drop fuzzy shadow,
  fonttitle=\bfseries\large,
  coltitle=white,
  colbacktitle=black!92,
  attach boxed title to top left={xshift=2mm,yshift=-2mm},
  boxed title style={
    colback=gray!80!black, % Or try a nice Navy: blue!70!black
    arc=1.5mm, 
    boxrule=0pt,
  },
  title={#2}, % Automatic numbering + Title
  list entry={#2}, % For the List of Instructions  
  #1
}

\usepackage{xspace}

\usepackage{pifont}

\newcommand{\cmark}{\textcolor{green!55!black}{\ding{51}}} % check
\newcommand{\xmark}{\textcolor{red!70!black}{\ding{55}}}   % cross
\newcommand{\pmark}{\textcolor{orange!90!black}{\raisebox{0.15ex}{\large$\blacktriangle$}}} % partial

\newcolumntype{L}{>{\raggedright\arraybackslash}p{0.31\linewidth}}
\newcolumntype{C}{>{\centering\arraybackslash}p{0.135\linewidth}}

\begin{document}

\title[A Unified Model for Search, Recommendation and Reasoning]{A Unified Language Model for Large Scale Search, Recommendation, and Reasoning}

\author{
    Marco De Nadai, Edoardo D'Amico, Max Lefarov\texorpdfstring{\textsuperscript{$\dagger$}}{1}, Alexandre Tamborrino, Divita Vohra, Mark VanMiddlesworth, Shawn Lin, Jacqueline Wood, Jan Stypka\texorpdfstring{\textsuperscript{$\dagger$}}{2}, Eliza Klyce, Keshi Dai, Timothy Christopher Heath, Martin D. Gould, Yves Raimond, Sandeep Ghael, Tony Jebara\texorpdfstring{\textsuperscript{$\dagger$}}{3}, Andreas Damianou, Vladan Radosavljevic, Paul N. Bennett, Mounia Lalmas and Praveen Chandar
}
\affiliation{%
  \institution{Spotify}
  \country{Denmark, Spain, Germany, France, United States, United Kingdom}
}
\email{mdenadai, edoardod, alexandret, divitav, mvanmiddlesworth, weihsiangl, jacquelinew, elizak, keshid, theath,}
\email{mg, yvesr, sandeepg, andreasd, vladanr, pbennett, mounial, praveenr@spotify.com}

\thanks{$\dagger$ Work performed while at Spotify.}
\renewcommand{\shortauthors}{De Nadai et al.}

\begin{abstract}
  Large language models (LLMs) are increasingly applied to recommendation, retrieval, and reasoning, yet deploying a single end-to-end model that can jointly support these behaviors over large, heterogeneous catalogs remains challenging. Such systems must generate unambiguous references to real items, handle multiple entity types, and operate under strict latency and reliability constraints requirements that are difficult to satisfy with text-only generation. While tool-augmented recommender systems address parts of this problem, they introduce orchestration complexity and limit end-to-end optimization.

We view this setting as an instance of a broader research problem: how to adapt LLMs to reason jointly over multiple-domain {\em entities}, user behavior, and language in a fully self-contained manner. To this end, we introduce {\bf NEO}, a framework that adapts a pre-trained decoder-only LLM into a tool-free, catalog-grounded generator. NEO represents items using {\em semantic identifiers (SIDs)} and trains a single model to \emph{interleave} natural language and typed item identifiers within a shared sequence. Natural-language prompts control the task, target entity type, and output format (IDs, text, or mixed), while constrained decoding guarantees catalog-valid item generation without restricting free-form text. We refer to this instruction-conditioned controllability as \emph{language-steerability}.

Inspired by multimodal alignment, we treat SIDs as a distinct modality and study design choices for integrating discrete entity representations into LLMs via staged alignment and instruction tuning. We evaluate NEO at scale on a real-world catalog of over 10M items across multiple media types and discovery tasks, including recommendation, search, and user understanding. In offline experiments, NEO consistently outperforms strong task-specific baselines and exhibits positive cross-task transfer, demonstrating a practical path toward consolidating large-scale discovery capabilities into a single language-steerable generative model.
\end{abstract}






\maketitle

\section{Introduction}
\begin{figure}[t]  
\begin{center}  
\includegraphics[width=\columnwidth]{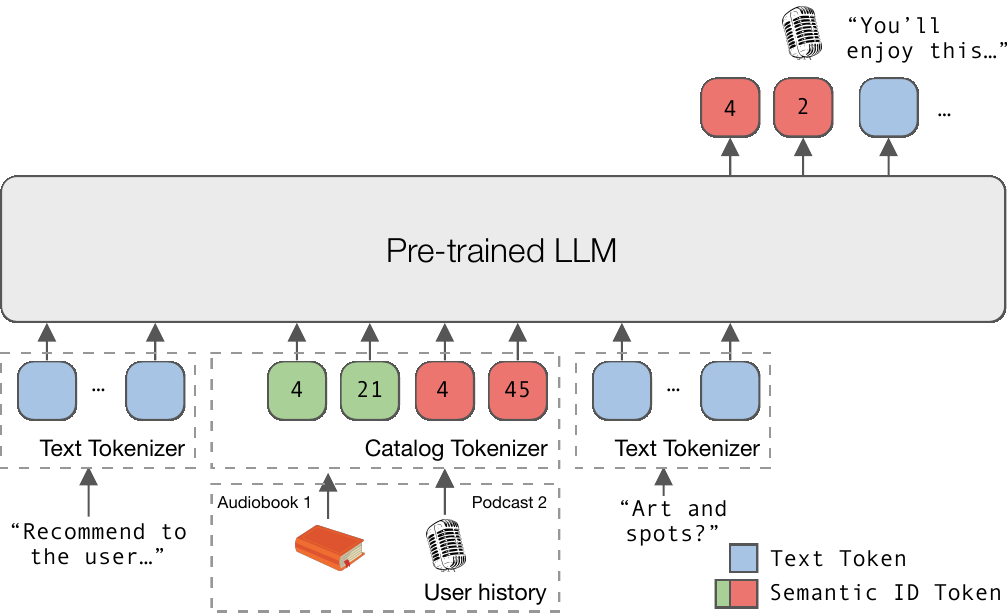}  
\end{center}  
  \vspace*{-0.1in}
\caption{NEO adapts pre-trained LLMs to \emph{speak} multi-item-type catalogs (e.g. audiobooks and podcasts) through SIDs (e.g. 2 tokens long tuples), and to output both text and SIDs.} \label{fig:teaser} 
  \vspace*{-0.1in}
\end{figure} 

Large-scale discovery systems operate over massive catalogs of discrete items (e.g., documents, media, products) and must support multiple access modes, most prominently recommendation and search. As user interfaces become more conversational, discovery interactions increasingly interleave explicit language-based queries with implicit behavioral feedback. Despite this convergence at the interaction level, recommendation and search are still typically implemented as separate system stacks~\cite{xu2018deep, penha2025semantic}, leading to duplicated modeling, fragmented objectives, and substantial operational overhead. While recent work explores unifying these capabilities through generative modeling~\cite{penha2025semantic, 10.1145/3640457.3688123}, achieving practical consolidation at scale, especially over heterogeneous catalogs with multiple entity types, remains challenging.

Large Language Models (LLMs) offer a natural abstraction for unifying discovery, providing a single sequence-modeling paradigm and a \emph{steerable} natural-language interface~\cite{brown2020language, ouyang2022training}
However, deploying a single LLM for large-scale discovery is difficult when model outputs must unambiguously reference real catalog items. Purely textual item mentions~\cite{2025_STARSimpleTrainingfreeApproachRecommendations, he2023large} are ambiguous and unstable, embedding-based item sequences~\cite{ning2025user} require architectural changes and struggle with long histories, and tool-augmented pipelines~\cite{zhao2024let, friedman2023leveraging, doh2025talkplay, wang2024recmind} introduce orchestration complexity that limits end-to-end optimization and latency that adds practical challenges
~\cite{schick2023toolformer, yao2022react,sharma2025retrieval}. In contrast, traditional generative recommenders~\cite{kang2018self, geng2022recommendation} have been demonstrated to scale on large datasets~\cite{zhai2024actions}, but they operate on sequences of atomic IDs and cannot interleave language with explicit item references or generate item-grounded explanations, particularly in heterogeneous catalogs.

Motivated by these limitations, we focus on a constrained but practically relevant setting: a massive, heterogeneous catalog in which a single deployed model must simultaneously support recommendation, text-based retrieval, and item-grounded text generation, while remaining \emph{steerable through natural-language instructions}, and operating end-to-end and tool-free at inference time.
We view this setting as an instance of a broader research problem: how to adapt LLMs to reason jointly over domain entities, user behavior, and language in a fully self-contained manner.

We introduce {\bf NEO}, a framework that adapts an open-weight decoder-only LLM into a single end-to-end, catalog-grounded generator. NEO represents items using semantic identifiers (SIDs) ~\cite{rajput2024recommender, singh2024better} and trains a single model to consume and produce sequences that interleave natural language with typed item identifiers (see \Cref{fig:teaser}). Natural-language prompts control the task, target entity type, and output format (IDs, text, or mixed), while constrained decoding guarantees that generated identifiers correspond to valid catalog items without restricting free-form text. Inspired by multimodal alignment, we treat SIDs as a distinct modality and integrate them via a staged training recipe comprising semantic foundation, domain grounding, and capability induction, while preserving the standard next-token prediction objective.

Within this framework, we examine a set of research questions relevant to general design choices in domain-adapted LLMs. We study the role of semantically structured identifiers versus atomic IDs, the impact of different alignment strategies on generalization and language retention, and the ability of a single model to support multiple discovery tasks under explicit language control. To ground these questions in practice, we evaluate NEO at scale on a catalog exceeding 10M items from a major streaming platform, spanning multiple item types and tasks including recommendation, search, and user understanding. Across offline experiments, NEO consistently outperforms strong task-specific baselines, exhibits positive cross-task transfer, and demonstrates a practical path toward consolidating large-scale discovery into a single language-steerable and catalog-grounded generative model.

\section{Related work}

We review related work through the lens of the broader problem introduced in the previous section: adapting LLMs to operate over domain entities and user behavior in a fully self-contained manner, while remaining language-steerable and scalable to large, heterogeneous catalogs. Prior work addresses aspects of this challenge across representation learning, generative recommendation, and LLM-based personalization, but typically resolves key requirements--such as item grounding, task control, language, or scalability--by externalizing complexity to specialized architectures, auxiliary tools, or task-specific pipelines. We organize related work according to how these design choices trade off generality, steerability, and end-to-end integration.

\begin{table*}[t]
\small
  \centering
  \caption{Comparison of NEO with prior recommendation paradigms. While several approaches support individual capabilities (e.g., discrete item IDs or LLM-based generation), NEO is the first to combine typed item addressing, mixed text–item outputs, language-steerable behavior, and multi-task discovery within a single tool-calling-free generative model.}
    \vspace*{-0.1in}
  \label{tab:killer_capabilities}
  \begin{tabularx}{\textwidth}{X C C C C C}
    \toprule
    \textbf{Capability $\downarrow$ / System $\rightarrow$} &
    \textbf{Generative SID} \textbf{Retrieval~\cite{2024_TokenRecLearningTokenizeIDLLMbaseda, rajput2024recommender}} &
    \textbf{Tool-based LLMs} \textbf{\cite{doh2025talkplaytools, friedman2023leveraging, lyu2024llm, wang2024recmind}} & \textbf{PLUM~\cite{2025_PLUMAdaptingPretrainedLanguageModels}} &\textbf{TalkPlay~\cite{doh2025talkplay}} &
    \textbf{NEO} \textbf{(this work)} \\
    \midrule
    Discrete item pointers           & \cmark & \xmark & \cmark & \cmark & \cmark\\
    Mixed text + item output           & \xmark & \pmark & \xmark & \cmark & \cmark\\
    Mixed text + item input             & \pmark & \cmark & \cmark & \cmark & \cmark\\
    User behaviour in input             & \cmark & \pmark & \cmark & \pmark & \cmark\\
    Language-steerable behavior        & \xmark & \cmark & \pmark & \cmark & \cmark\\
    Multiple item types               & \pmark & \pmark & \xmark & \xmark & \cmark\\
    Search + Rec + Explanation         & \xmark & \pmark & \xmark & \xmark & \cmark\\
    Tool-calling-free inference               & \cmark & \xmark & \cmark & \pmark & \cmark\\
    %Single model, single pass         & \xmark & \xmark & \xmark & \xmark & \cmark \\
    Low-latency &  \cmark & \xmark & \cmark & \cmark & \cmark\\
    Industrial-scale (>10M items)    & \cmark & \xmark & \cmark & \xmark & \cmark\\
    \bottomrule
  \end{tabularx}

  %\vspace{3pt}
  {\footnotesize
  \textbf{Legend:} \cmark\ explicitly supported and demonstrated;\;
  \pmark\ partially supported / requires extra components;\;
  \xmark\ not supported / not demonstrated / not studied.}

\end{table*}

%We review prior work on recommender systems, large language models, and item-grounded generation, focusing on approaches to unifying language and item representations at scale.

%\textbf{Generative models for recommendations.}
\paragraph{Generative models for recommendations.}
Generative recommendation models address the entity grounding problem by predicting items autoregressively, but often assume a single task and a homogeneous item space (e.g., next-item prediction)~\cite{kang2018self, zhai2024actions, li2025realizing}.
%A central challenge in generative recommendation models (e.g.~\cite{kang2018self}) is item representation. 
Many approaches rely on atomic item IDs, which limits generalization in cold-start and long-tail regimes~\cite{2024_BetterGeneralizationSemanticIDsCase}. 
These models are typically designed for single-task next-item prediction and operate purely over ID sequences, making it difficult to incorporate natural language and support multiple discovery tasks within a single model. 

To address limitations of atomic IDs, recent work has explored both continuous item embeddings~\cite{ning2025user} and SIDs~\cite{singh2024better, rajput2024recommender}. Although continuous embeddings are effective in many multimodal settings (e.g., \cite{liu2023visual}), injecting long sequences of dense vectors into an LLM can be bandwidth- and memory-intensive for users with long interaction histories, and often requires architectural modifications (e.g., projection layers or adapters) or additional training objectives beyond standard next-token prediction. In contrast, SIDs can be treated as ordinary tokens by both traditional and generative sequence models, while preserving the semantic neighborhood structure that is richer than atomic item IDs~\cite{singh2024better}.

One influential line of work that operationalizes SIDs for recommendation begins with TIGER~\cite{2024_BetterGeneralizationSemanticIDsCase}. TIGER introduces a pipeline that learns item embeddings, discretizes them into SIDs, and applies autoregressive modeling for recommendation. Subsequent work improves identifier quality using retrieval-oriented or contrastive objectives~\cite{2024_CoSTContrastiveQuantizationbasedSemantic, 2024_STOREStreamliningSemanticTokenizationGenerative}, as well as by incorporating collaborative signals during code learning~\cite{2024_AdaptingLargeLanguageModelsIntegrating, 2023_CoLLMIntegratingCollaborativeEmbeddingsLarge, 10.1145/3705328.3759300}. However, most SID-based recommenders are not language-steerable and do not support mixed text-ID inputs or outputs, limiting their applicability to tasks such as search and natural-language explanation. In addition, their interfaces offer limited flexibility for rich textual inputs (often relying on soft tokens~\cite{ning2025user, huber2025embedding}) and do not naturally support operating over multiple item types within a single model. We summarize these capability gaps in \Cref{tab:killer_capabilities}.
From the perspective of domain-adapted LLMs, these approaches provide strong item-level modeling but lack a unified interface for language steerability, multi-task discovery, and heterogeneous entity types.

%\textbf{LLMs for personalization and discovery.}
\paragraph{LLMs for personalization and discovery.}
Recent work applies LLMs to personalization by introducing natural-language interfaces for recommendation and discovery, framing user intent and context directly in text~\cite{2025_STARSimpleTrainingfreeApproachRecommendations,lyu2024llm,xu2023openp5}.
These approaches rely on text-only representations~\cite{2025_STARSimpleTrainingfreeApproachRecommendations, he2023large}, textual and rule-based IDs~\cite{2024_IDGenRecLLMRecSysAlignmentTextualID}, or atomic  IDs~\cite{xu2023openp5, doh2025talkplaytools, zhang2025recommendation}, and some explore unifying search and recommendation within a single model~\cite{10.1145/3705328.3759300, 10.1145/3640457.3688123}. However, such methods are often evaluated on relatively small single-typed datasets, struggle with cold-start items when SIDs are not used, or face scalability limitations when applied to large catalogs, particularly when relying on free-form textual references to items.

To address item ambiguity and expand capability, several systems augment LLMs with external tools or retrievers for personalized recommendation~\cite{friedman2023leveraging, doh2025talkplay, wang2024recmind, lyu2024llm}. For example, TalkPlay-tools~\cite{doh2025talkplaytools} orchestrates multiple tools via planning to answer user queries, while RecMind~\cite{wang2024recmind} relies on external knowledge and tool use for zero-shot personalization. While effective in certain settings, these approaches are not trained end-to-end, operate primarily in the text modality, leading to ambiguity when referencing catalog items, and introduce additional orchestration complexity and latency at inference time. We summarize these trade-offs in \Cref{tab:killer_capabilities}.

\paragraph{SIDs as new modality in LLMs.}
% add two early works
  SIDs offer a discrete representation of entities that preserves semantic structure while remaining compatible with autoregressive generation~\cite{rajput2024recommender, singh2024better}.
 %Recent work has begun exploring the use of SIDs within LLM-based recommender systems. 
 TalkPlay~\cite{doh2025talkplay} reformulates music recommendation as token generation by expanding the LLM vocabulary with feature-engineered multimodal music tokens, enabling single turn interactions that interleave text and SIDs in both input and output. However, this approach focuses on a single item type and task, relies on an external retriever operating over SIDs, is not evaluated at large scale, and does not incorporate user history for personalization.

Concurrent work such as PLUM~\cite{2025_PLUMAdaptingPretrainedLanguageModels} explores a related direction based on continual pre-training and supervised fine-tuning with SIDs, and shows partial language-steerability through qualitative in-context learning examples. However, PLUM currently does not support mixed text–ID outputs in downstream tasks, focuses exclusively on recommendations, and is limited to a single item type.
Prior SID-based systems primarily target single-task recommendation or retrieval, and do not study how discrete entity representations can be integrated into a language-steerable, multi-task LLM that supports mixed text+item input {\em and} output while maintaining general language competence.

\paragraph{NEO} In contrast to prior work that applies SIDs within LLMs primarily for recommendation, we introduce a typed mixed-sequence catalog interface and a staged alignment recipe. This enables an open-weight decoder-only LLM to interleave natural language with valid identifiers from a heterogeneous catalog and support recommendation, text retrieval, and item-grounded text generation and user understanding in a single, tool-free model.

% ==================
\section{Methodology}
\begin{figure*}[ht!]  
\begin{center}  
\includegraphics[width=\textwidth]{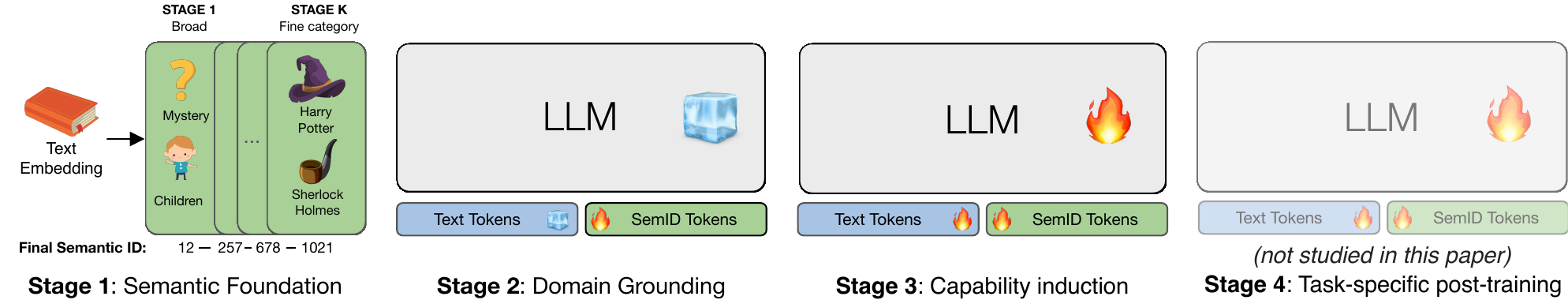}  
\end{center}  

\caption{We illustrate our approach as the first three stages in a more general four-stage pipeline. In the general case, in Stage~1, semantic representations are learned---such as semantic IDs (SIDs)---to capture entities, their structure, and relationships. % (e.g. collaborative-filtering contrastive learning). 
In Stage~2, domain grounding is performed via one or both of (i) adaptation over user-behavior sequences and (ii) domain alignment between SID/text pairs, to learn a shared latent space between entities and natural language. This enables the model to use text and SIDs interchangeably and to learn sequential, domain-relevant concepts. Stage~3 teaches the model to answer questions, follow commands, and generate grounded outputs through multi-task instruction-following. We prioritize diversity of tasks over depth within any single task. In Stage~4, we may leverage task-specific datasets and objectives related to the targeted use case (e.g., topic discovery and business constraints) via task-specific fine-tuning, reinforcement learning, and related techniques. We indicate typical choices for which parameters are kept frozen (ice) or trainable (fire) for each stage.}
\label{fig:architecture_full_stack} 
\end{figure*} 

Figure~\ref{fig:architecture_full_stack} presents a general staged pipeline for training a domain-adapted, fully self-contained LLM, of which NEO instantiates the first three stages. Stage~1 learns SIDs as discrete token sequences representing catalog entities and their semantic structure. Stage~2 grounds these identifiers by aligning SID-text pairs (and optionally adapting over user-behavior sequences) to learn a shared space where text and SIDs are interchangeable, supporting mixed text–SID sequences and sequential domain signals. Stage~3 then induces discovery capabilities via multi-task instruction-following, prioritizing breadth over single-task specialization, whereas an optional Stage~4 captures subsequent task-specific objective optimization using larger datasets and specialized objectives (e.g., business-constraint and product-specific tuning, or reinforcement learning).
We compare with alternative training strategies, including continuous pre-training (CPT)-based approaches, in Section~\ref{sec:design_analysis}.

NEO adapts a pre-trained open-weight decoder-only LLM to operate jointly over natural language and discrete item identifiers, while preserving the standard next-token prediction objective and requiring no architectural modifications (e.g., no bespoke attention masks, cross-attention modules, or specialized losses). Text inputs are tokenized using the native tokenizer of the backbone LLM, while catalog items are represented as short sequences of SID tokens obtained by discretizing item embeddings. Both text tokens and SID tokens are embedded into a single shared sequence and processed by the same decoder stack, enabling the model to reason uniformly over language and items.

To reliably integrate this extended vocabulary and align item identifiers with linguistic representations, NEO is trained using a three-stage recipe, as shown in \Cref{fig:architecture_full_stack}. First, item content is converted into SIDs through embedding discretization (Stage~1). Second, the newly introduced SID tokens are aligned with the LLM's embedding space (Stage~2). Third, the model is instruction-tuned on a mixture of discovery tasks, including recommendation, retrieval, recommendation explanations, and user interest analysis (Stage~3).

\subsection{Semantic IDs}
In Stage 1, we focus on constructing discrete item representations that can be used directly by an LLM. We represent each catalog item $e$ with $\mathrm{SID}(e)$: a short sequence of $M$ discrete tokens, each with $K$ possible values, that compactly encodes the semantic content of the item. Intuitively, SIDs provide a discrete token-level abstraction of items that preserves semantic neighborhood structure while remaining compatible with autoregressive generation. See Appendix \ref{appendix:semid_formal} for formal details.

We construct SIDs by discretizing content-based item embeddings in a hierarchical fashion using residual quantizers (e.g.,~\cite{ju2025generative, lee2022autoregressive}), inducing a coarse-to-fine decomposition: earlier codes capture broad semantic regions, and later codes refine within-region distinctions. This property aligns naturally with autoregressive generation, where tokens are produced sequentially.
We design the discretization to ensure semantic coherence, stability under distribution shift (e.g., seasonality of content consumption and shifts in interests), and compatibility with large heterogeneous catalogs.

For heterogeneous entities, embeddings of different item types may lie in different latent spaces, thus requiring multiple quantizers.

\paragraph{Using SIDs in the LLM}
To integrate SIDs and natural language, we define a unified vocabulary $\mathcal{V}=\mathcal{V}_{\text{text}} \cup \mathcal{V}_{\text{SID}}$. We extend the LLM's base vocabulary by $M \times K$ SID tokens, which are randomly initialized, and add the corresponding parameters to the LLM head when their weights are untied.
SIDs are interleaved with text in both inputs and outputs. Following delimiter-based conventions commonly used in Multimodal LLMs (MLLMs) (e.g.,~\cite{2024_MIOFoundationModelMultimodalTokens, _VideoPoetLargeLanguageModelZeroShota, nguyen2025spirit}) item references are emitted only within explicit item spans: $\langle \text{text} \rangle\ \texttt{[SID]}\ \langle c_1\ \ldots\ c_M \rangle\ \texttt{[/SID]}\ \langle \text{text} \rangle$. This design allows the model to freely interleave natural language and item references while enabling constrained decoding within SID spans.

\subsection{Domain Grounding: Alignment}
Stage~2 aligns SIDs with the LLM's linguistic embedding space. This stage is inspired by MLLMs, where non-textual modalities are represented either as discrete tokens~\cite{li2025onecat, 2024_MIOFoundationModelMultimodalTokens, wu2026liquid, wang2024emu3, wu2024vila, nguyen2025spirit, team2024chameleon} or as continuous features interfaced through encoders and projection layers~\cite{wu2025janus, wang2024qwen2, sun2023emu}. 
MLLMs are typically trained using staged recipes that introduce a new modality, align it with language, and then apply instruction tuning. 
Following this paradigm, we treat SIDs as a discrete ``foreign-language'' modality and learn a bidirectional mapping between SID sequences and natural language. This enables the model to both \emph{consume} and \emph{generate} mixed text--SID sequences, while preserving the standard next-token prediction objective.

\paragraph{Bidirectional alignment objectives.}
Let $s_e=\mathrm{SID}(e)$ denote the SID sequence for item $e$, let $t_e$ denote a short textual descriptor of $e$ derived from available metadata, and let $c_e$ denote the item type of $e$. We train the model using three complementary  objectives:
\begin{enumerate}
    \item \textbf{SID$\rightarrow$text (verbalization).} Given one or more SID spans (optionally with a lightweight instruction), the model predicts a natural-language descriptor: $\langle \mathrm{prompt}(s_e)\rangle \rightarrow \langle t_e \rangle$
    \item \textbf{text$\rightarrow$SID (grounded retrieval).} Given a natural-language query $q$, the model predicts the corresponding item SID span: $\langle \mathrm{prompt}(q)\rangle \rightarrow \texttt{[SID]}\ \langle s_e\rangle\ \texttt{[/SID]}$. To reduce ambiguity and noise in the supervision, queries are constructed from a subset of item fields (primarily name and short description). We hypothesize that the more specific the prompts are, the better the alignment quality is, due to reduced semantic ambiguity.
    \item \textbf{SID$\rightarrow$type (type disambiguation).} Given one or more SID spans (optionally with a lightweight instruction), the model predicts the item type: $\langle \mathrm{prompt}(s_e)\rangle \rightarrow \langle c_e \rangle$.
\end{enumerate}

Together, these objectives encourage the model to align SIDs with language at the levels of semantics, retrieval, and type structure.

\paragraph{Parameterization and stability.}
To prevent catastrophic forgetting during the alignment phase, we freeze the pre-trained backbone parameters and optimize only: (i) the input embeddings corresponding to newly introduced SID tokens, and (ii) the output weights used to predict SID tokens (the SID-specific logits in the LLM head). This design preserves the linguistic competence of the backbone while enabling effective alignment between SIDs and natural language.

\subsection{Capability Induction via Instruction tuning}
Stage~3 trains the domain-grounded model to perform personalization tasks and discovery under explicit control via natural-language instructions, while producing (i) SIDs, (ii) text, or (iii) mixed SIDs--text outputs. We unfreeze and fine-tune all model parameters. Alternatively, Stage~3 can be implemented using parameter-efficient fine-tuning methods (e.g., LoRA~\cite{hu2022lora}) when full fine-tuning is impractical. In this paper, we focus on simply introducing a large variety of types of tasks to the model including text-based retrieval, recommendation, and text generation across heterogeneous items. We save for future work demonstrating the ability to generalize in this stage to new, zero-shot \emph{domain} capabilities.

We train NEO using supervised instruction tuning over the unified vocabulary $\mathcal{V}$, where each training example is a prefix–completion pair  $(x,y)$. The prefix $x$ is the input context and includes the system prompt and user instruction, optionally augmented with lightweight user features and/or item references expressed as SID spans. The completion $y$ is the model output and may contain natural language, SID spans, or both. During training, we apply the loss only to tokens in $y$, masking  all tokens in~$x$. 
We restrict $x$ to single-turn interactions to simplify evaluation; the same formulation extends directly to multi-turn dialogues by concatenating the full conversation history ($(x,y)$ pairs) and masking all user messages.

\paragraph{Task families}
\label{sec:tasks}
All tasks are described through prompts that follow a consistent schema that explicitly specifies (i) the task (recommend, retrieve,  explain, understand), (ii) the target item type, and (iii) the desired output format (SID-only, text-only, or mixed). This explicit conditioning enables fine-grained control over model behavior using natural-language instructions.

The tasks are formulated as next-token prediction over mixed text--SID sequences that may involve items from multiple types:
\begin{itemize}
    \item \textbf{Next-item recommendation.} Given an instruction, lightweight user or context features, and a short interaction history represented as SIDs, the model predicts the next consumed item of the specified type as a SID.
    \item \textbf{Text-based retrieval.} Given a natural-language query (optionally with lightweight user context), the model generates the SID(s) of the most relevant item(s).
    \item \textbf{Grounded mixed generation (recommendation + recs-planation).} The model generates a recommended item SID together with a natural-language justification grounded in the user's history and the recommended item(s).
    \item \textbf{User understanding (interest profiling).} Given a user's interaction history expressed as SIDs, the model generates a short natural-language description of the user's interests.
\end{itemize}

In settings where natural-language data is unavailable (e.g., user interest profiles), we prompt a stronger LLM to distill synthetic supervision for NEO’s training.

\subsection{Inference}
\label{sec:inference}
Because multi-token SIDs induce a large combinatorial output space (e.g., a 4-token SID with $K{=}256$ yields $256^4$ possible sequences), we constrain decoding so that generated SIDs always map to existing catalog items. We precompute all valid SID tuples and store them in a prefix trie. During autoregressive decoding, we apply trie-based masking only within \texttt{[SID]}...\texttt{[/SID]} spans, restricting each next token to trie-consistent continuations. Outside SID spans, we restrict generation to text tokens (and \texttt{[SID]} token), which prevents accidental SID token emission in free-form text while preserving the ability to introduce item references when needed.

The mapping from items to SID tuples is not necessarily one-to-one, as multiple items may share the same SID. We resolve such collisions by mapping each SID tuple to a single canonical item, selected based on multiple possible heuristics (e.g. item popularity, recency).
For retrieval tasks, we generate multiple candidate items by applying beam search over SID tokens.

\section{Evaluation}
We evaluate NEO at scale across recommendation, retrieval, and user understanding tasks, focusing on effectiveness, steerability, and robustness under realistic production constraints.

\subsection{Experimental setup.} 

We evaluate NEO using a sample of interaction logs from a large-scale media streaming platform, spanning a catalog of over 10M items and approximately 15M users. The catalog includes multiple item types, including episodes, shows,\footnote{We define a show as a podcast series, composed of episodes.} audiobooks, and artists. We use around 5M examples in the domain grounding stage, while in the capability induction stage we train on a total of 10M training examples and test on around 100K examples.

\subsubsection{Baselines} We compare the NEO-trained LLM against strong and mature production systems optimized over multiple years using large-scale data and extensive feature engineering. 

For recommendation tasks (audiobooks, shows, and episodes), the baseline combines Graph Neural Networks with a two-tower architecture~\cite{yi2019sampling} and models cross-item-type relationships through item co-consumption patterns, inspired by~\cite{10.1145/3589335.3648339}. It incorporates weak signals (e.g., follows and previews prior to streams) to better handle cold-start users, and uses categorical features (e.g., categories, topics, user country) alongside LLM-encoded item metadata.

For text-based retrieval, the baseline is a dense retrieval system similar to~\cite{yang2021universal}, trained on historical search logs. Training data includes query-entity pairs, multi-step reformulation sessions (in which the user had to reformulate the query to find the desired content), manually curated semantic queries, and synthetic queries generated as done in~\cite{liang2020embedding}. The query encoder processes a textual concatenation of the user’s geographic location and the search query, while the entity encoder represents item metadata using concatenated metadata such as name and description.

We evaluate recommendation and text-based retrieval performance using standard ranking metrics: Hit Rate@K (HR@K) and Normalized Discounted Cumulative Gain@K (NDCG@K), with $K \in \{10, 30\}$ (see \Cref{sec:metrics} for  details). 

There are no baselines for grounded text-SID mixed generation. These tasks are completely novel in our setting. 

\subsubsection{Semantic ID creation (Stage 1)} 
\label{sec:our_stage1}
Each item is represented using its available content and metadata. For artists, we encode tracks from track spectrograms into audio embeddings and average them by artist to obtain artist-level vectors.
For shows, episodes, and audiobooks, we embed title and description text using the Qwen3 embeddings (8B) pre-trained text model~\cite{zhang2025qwen3}. 
Metadata quality can vary, particularly for podcast episodes, where descriptions may be missing or largely identical across episodes of the same show. To mitigate this, we generate richer episode summaries using a pre-trained LLM applied to multiple available metadata fields, and embed these summaries using the same text encoder.

We construct SIDs using Residual $K$-means~\cite{ju2025generative} with $M{=}4$ codebooks. We use $K{=}256$ centroids per codebook for audiobooks, episodes and shows, and $K{=}1024$ for artists to better model the higher complexity of audio embeddings. We construct separate SID vocabularies per item type, as early experiments showed that embedding spaces differ substantially across types.
All SID tokens, along with span delimiters \texttt{[SID]} and \texttt{[/SID]} are added to the LLM vocabulary, extending the embedding layer and LLM head by $7168 + 2$. To mitigate the curse of dimensionality of $K$-means, we reduce embedding dimensionality to $1024$ dimensions using the Matryoshka property~\cite{kusupati2022matryoshka} of Qwen3 embeddings.
We solve collisions with a popularity-based selection for items colliding in the same SID bucket. \Cref{sec:semantic_foundation_eval} challenges this solution without seeing significant differences from random selection.

In our heterogeneous setting, using a single quantizer for all item embedding spaces would severely compromise SIDs quality.

\subsubsection{Domain Alignment (Stage 2)}
We construct paired alignment examples linking each item SID $s_e$ with a short textual descriptor $t_e$ derived from metadata. 
For audiobooks, episodes, and shows, descriptors include titles, descriptions, summaries, topics, and categories. For artists, descriptors include names, genres, and Wikipedia snippets when available. Text-to-SID queries $q$ are constructed from a subset of these fields to reduce ambiguity.

\subsubsection{Capability Induction via Instruction Tuning (Stage 3)}

This stage fine-tunes the aligned model on multiple supervised instruction-following discovery tasks. All prompts use a consistent instruction-tuning format specifying task, target item type, and output constraints (SID-only, text-only, or mixed). Prompt templates are provided in \Cref{appendix:prompts}.

\begin{description}[font=\normalfont\itshape, leftmargin=1em]
    \item[Next-item(s) recommendation.] Each example consists of an instruction, user context, and target label. User context includes recent interaction history represented as SIDs. For audiobooks and shows, we include recent cross-item types activity, following~\cite{10.1145/3589335.3648339}.

\item[Text-based retrieval.]
We build a dataset mapping natural-language queries to the SID of the most relevant item of a specified type, using engagement logs as supervision. Prompts may optionally include lightweight user context.

\item[User understanding (interest profiling).]
Explicit natural-language supervision for user interest profiling is typically unavailable. We therefore create training targets via distillation from a larger LLM (32B parameters), which is prompted with textual descriptions of a user’s consumed items to generate a concise interest summary. We then replace textual item representations with SIDs and train NEO to generate the same summary directly from the SID-based history, without intermediate chain-of-thought.

\item[Recommendation + explanation (``recsplanation'').] We build a mixed output dataset where the target consists of (i) a recommended item SID and (ii) a natural-language justification grounded in the user’s history and the recommended item. Rationales are obtained via distillation from a larger LLM: for each user sequence, we sample a cut point $t$ and prompt the LLM to explain why the item at $t+1$ matches the user’s prior interests. We then replace textual item mentions with SIDs and train NEO to reproduce the rationale while outputting the correct SID.

\end{description}

%We construct a mixed-output dataset where the target consists of (i) a recommended item SID and (ii) a natural-language justification grounded in the user history and the recommended item. We obtain natural-language rationales via teacher distillation: for each user sequence, we sample a cut point $t$ and prompt the teacher to explain why the item at $t{+}1$ matches the user's prior interests. We then swap the textual item representations in the prompt with SIDs and train NEO to reproduce the rationale while outputting the correct SID.

\subsubsection{Implementation details} 
\label{sec:temporal_split}
Unless otherwise stated, we use Qwen3-0.6B~\cite{yang2025qwen3} as decoder-only backbone. To evaluate backbone generality, we also report results with Llama 3.2 1B~\cite{grattafiori2024llama}. 
We follow a global temporal evaluation protocol~\cite{meng2020exploring}. For each example, the context contains interactions up to day $t$ and the label corresponds to the item consumed at $t{+}k$ with $k{=}1$ for episodes and shows and $k{=}7$ for audiobooks. We prevent temporal leakage across all tasks and evaluate at $t+2k$.
To ensure fair comparison, NEO and all baselines are trained on the same temporal window and evaluated on an identical held-out test window under the same candidate universe. 

For mixed text-SID generation tasks, we build ~ 20 types of prompt templates that are randomly selected for each data sample and filled with user information, to enhance the diversity of our examples.

\paragraph{Cold-start users.} We simulate cold-start users by removing past listening behavior information and rely only on user metadata for predictions on recommendations and text-based retrieval.

%\subsection{Empirical Analysis of Item Tokenization, Alignment, and Decoding}
\subsection{Design Analysis Research Questions \label{sec:design_analysis}}

We analyze the design choices underlying NEO’s item tokenization, alignment, and decoding as it relates to the general stages we present for fully self-contained domain-adapted LLMs in Figure \ref{fig:architecture_full_stack}.

\subsubsection{Semantic Foundation}
\label{sec:semantic_foundation_eval}
We examine generalization effectiveness of SIDs versus atomic IDs, importance of semantic structure in the SID token space, importance of the grounding content quality, whether learning user preferences among entities, from Collaborative Filtering (CF) data, is critical at this stage, whether constrained decoding is necessary, and its practical scale implications. 

\begin{table}[htb]
\small
  \centering
  %\resizebox{\textwidth}{!}{ 
  \caption{Ablation for multiple types of item representation.}
    \vspace*{-0.1in}
  \label{tab:ablation:semid}
  \begin{tabular}{l|c}
   \toprule[1.25pt]
    \textbf{Method}
    
      & \textbf{$\Delta$ HR@10}\\
    \midrule[1.1pt]
    \textcolor{gray}{\texttt{NEO (Baseline)}} & \textcolor{gray}{-}
    \\
    A - Atomic item IDs
    & \textcolor{scorered}{-59.7\%}
    \\
    \rowcolor[rgb]{0.93,0.93,0.93}
    \multicolumn{2}{c}{\texttt{Semantic IDs}} \\
    B - LSH quantizer  
    & \textcolor{scorered}{-51.2\%}
      \\
    C - No augmented data
    & \textcolor{scorered}{-2.9\%}
    \\
    % \texttt{LLM} - unique space  
    % & 
    % \\
    D - Collaborative filtering
    & \textcolor{scorered}{-25.6\%}
      \\
    \bottomrule[1.25pt]
  \end{tabular}%}
    \vspace*{-0.1in}
\end{table}

\paragraph{Do semantically-structured discrete item tokens improve tool-free generation over large catalogs beyond atomic IDs?}  To isolate the impact of semantic structure from the use of short discrete codes, we construct an ``atomic ID'' baseline by randomly permuting the SID tuples across items. This preserves the same code length and vocabulary size as NEO, but removes semantic neighborhood structure (items that are close in embedding space are no longer assigned similar codes). Because our SID configuration supports up to $K^M \approx 4$B distinct tuples, this provides a practical approximation of unique atomic identifiers at our catalog scale.

\Cref{tab:ablation:semid}~(A) shows that semantically structured item representations  consistently improve performance. Compared to purely ID-based modeling, SIDs enable better generalization across related episodes and reduce sensitivity to sparsity and cold-start items. These findings align with prior work~\cite{rajput2024recommender, singh2024better} in settings involving a single item type and a single task.

\paragraph{Does how the discretize embeddings support semantic structure matter (learned vs.\ random partitions)?}
To isolate the effect of the discretization method from that of the underlying embedding semantics, we compare  Residual $K$-means SIDs to Locality-Sensitive Hashing (LSH)~\cite{charikar2002similarity} SIDs. Both approaches operate on the same semantic space, but differ in how embeddings are mapped to discrete codes: Residual $K$-means learns codebooks from data, whereas LSH uses data-independent random projections that preserve similarity only in expectation. 
\Cref{tab:ablation:semid}~(B) shows a clear performance drop when using LSH-based SIDs, indicating that learned, distribution-adaptive quantization produces more informative SID token neighborhoods than data-independent hashing in our setting.

\paragraph{How sensitive are SIDs to metadata quality?} 
Because the SID construction is driven by the geometry of item embeddings, we evaluate sensitivity to metadata quality by comparing the SIDs created from raw metadata vs.\ those created from augmented, cleaner metadata (\Cref{sec:our_stage1}).
\Cref{tab:ablation:semid}~(C) shows that the use of raw metadata leads to a 2.91\% drop in performance. We attribute this degradation to lower-quality content embeddings caused by missing, one-liner text, or repetitive descriptions, which increase collisions and reduce semantic separability in the resulting SID space.

\paragraph{When do content-based vs.\ collaborative-based SIDs work better?} 
Introducing CF-based signals in SIDs is a way of learning relationships among entities beyond content, using user behavioural signals directly from the Semantic Foundation stage. Motivated by prior work suggesting that CF signals can improve discrete tokenization~\cite{2025_EAGERLLMEnhancingLargeLanguageModels}, we evaluate whether CF-derived representations consistently yield better SIDs than content-based embeddings. We construct CF-based representations for artists by training a word2vec-style model~\cite{mikolov2013efficient} on artist co-occurrence in playlists.
Despite performing strongly in traditional production recommender systems, CF-derived artist SIDs perform significantly worse than content-derived artist SIDs within our framework (\Cref{tab:ablation:semid}~(D)). This indicates that, in our setting, CF vectorization produce inferior semantic identifiers for generative, language-interleaved modeling. Overall, CF-derived SIDs underperform content-based SIDs and exhibit instability over time (see Appendix \ref{appendix:jaccard}), complicating alignment across item types.

\begin{table}[htb]
\small
  \centering
  \caption{Accuracy/latency trade-off at inference time.}
  \vspace*{-0.1in}
  \label{tab:beam_search}
  %\resizebox{\textwidth}{!}{ 
  \begin{tabular}{l|cc|c}
    \toprule[1.25pt]
    \multirow{2}{*}{\textbf{Inference strategy}} 
      & \multicolumn{2}{c|}{\textbf{$\Delta$ HR@K}} 
      & \textbf{$\Delta$} latency \\
    \cmidrule(lr){2-3}\cmidrule(lr){4-4}
      & \textbf{$k=10$} & \textbf{$k=30$} & \textbf{$k=10$}  \\
    \midrule[1.1pt]
    \textcolor{gray}{\texttt{NEO (Baseline)}} & \textcolor{gray}{-}& \textcolor{gray}{-}& \textcolor{gray}{-}
    \\
    A - Beam Search w/o constrains & \textcolor{scorered}{-1.80\%} & \textcolor{scorered}{-1.93\%} & \textcolor{scoregreen}{-2.92\%}
    \\
    B - Top-$p$ sampling & \textcolor{scorered}{-21.97\%} & \textcolor{scorered}{-32.41\%} & \textcolor{scoregreen}{-7.91\%}
      \\
    \bottomrule[1.25pt]
  \end{tabular}%}
    \vspace*{-0.1in}
\end{table}
\paragraph{What is the accuracy-latency trade-off of constrained decoding for catalog-valid generation?} 
As SID tuples induce a large combinatorial output space, we evaluate whether catalog-validity constraints (implemented via prefix-trie-constrained decoding) are necessary for accuracy.
\Cref{tab:beam_search}~(A) shows that, even without constrained decoding, the model generates valid SIDs in ~98\% of the case. 
Adding catalog-validity constraints incurs only a small latency overhead, while enabling greater flexibility at inference time (e.g., restricting generation to a subset of items such as newly added content).

We also compare Beam search with 30 beams against top-$p$ sampling, using Qwen's default decoding hyperparameters (temperature $=0.6$, top-$k=20$, top-$p=0.95$). \Cref{tab:beam_search}~(B) shows that the performance degradation induced by Beam Search is probably justified by the significant lower accuracy of Top-$p$ sampling.

Finally, we evaluate the sensitivity of inference-time performance to the popularity-based collision resolution described in \Cref{sec:inference} by randomly selecting an item within each collision bucket. We observe no significant performance differences, indicating that SID collisions are rare at inference time and do not materially affect results.

\subsubsection{Domain Grounding}
Here, we focus on alignment and examine whether a staged training recipe is necessary for effective generalization, how different optimization strategies (e.g., CPT vs.\ our Alignment) affect retention of language capabilities, and whether the approach truly generalizes across different LLM backbones.

\begin{table}[htb]
\small
  \centering
  \caption{Comparison of training strategies alternative to NEO.}
    \vspace*{-0.1in}
  \label{tab:ablation:training}
  \setlength{\tabcolsep}{4pt}
  %\resizebox{\textwidth}{!}{ 
  \begin{tabular}{l|cc|cc}
    \toprule[1.25pt]
    \multirow{2}{*}{\textbf{Method}} 
      & \multicolumn{2}{c|}{\textbf{HR@K$\uparrow$}} 
      & \multicolumn{2}{c|}{\textbf{NDCG@K$\uparrow$}} \\
    \cmidrule(lr){2-3}\cmidrule(lr){4-5}
      & \textbf{$k=10$} & \textbf{$k=30$} & \textbf{$k=10$} & \textbf{$k=30$} \\
    \midrule[1.1pt]
    \textcolor{gray}{\texttt{NEO (Baseline)}} 
    & -
    & -
    & -
    & -
    \\
    A - w/o Domain Grounding
    & $\textcolor{scorered}{-6\%}$
    & $\textcolor{scorered}{-7\%}$
    & $\textcolor{scorered}{-8\%}$
    & $\textcolor{scorered}{-8\%}$
    \\
    B - Everything in two-stages
    & $\textcolor{scorered}{-8\%}$
    & $\textcolor{scorered}{-7\%}$
    & $\textcolor{scorered}{-10\%}$
    & $\textcolor{scorered}{-10\%}$
    \\
    C - Random init
    & $\textcolor{scorered}{-44\%}$
    & $\textcolor{scorered}{-44\%}$
    & $\textcolor{scorered}{-44\%}$
    & $\textcolor{scorered}{-43\%}$
    \\
    D - Continuos pretraining
    & $\textcolor{scorered}{-3\%}$
    & $\textcolor{scorered}{-3\%}$
    & $\textcolor{scorered}{-2\%}$
    & $\textcolor{scorered}{-3\%}$
    \\
    \bottomrule[1.25pt]
  \end{tabular}%}
    \vspace*{-0.1in}
\end{table}

\paragraph{Is a staged alignment recipe necessary to integrate SID tokens into an LLM without degrading language competence?}
We evaluate whether NEO’s staged training is necessary by comparing it to: (i) removing the Domain Grounding stage, (ii) collapsing Domain Grounding and Capability Induction into one stage, and (iii) randomly initialized parameters without a pre-trained backbone.

\Cref{tab:ablation:training}~(A) shows that the three-stage recipe is critical for both effectiveness and optimization. Removing the Domain Grounding stage consistently degrades retrieval performance and slows convergence under the same compute budget. Collapsing Stage 2 and 3 (\Cref{tab:ablation:training}~(B)) further reduces performance, consistent with increased objective interference and weaker grounding of SID tokens. (\Cref{tab:ablation:training}~(C)) Training from scratch performs substantially worse, highlighting the importance of pre-trained language representations for sample-efficient learning and strong final accuracy.

\paragraph{Using our Alignment vs Continuous pretraining (CPT) in Domain Grounding}
Table \ref{tab:ablation:training} (D) shows that the CPT approach used in PLUM~\cite{2025_PLUMAdaptingPretrainedLanguageModels} slightly underperforms NEO-Qwen on downstream tasks. However, CPT causes a severe drop in general knowledge; its MMLU-Redux~\cite{gema2024mmlu} score plunges to 0.03, whereas NEO-Qwen maintains the original backbone score of 0.46 after Domain Grounding.
 
This difference matters because NEO targets tasks that require robust instruction following and grounded natural-language generation (e.g., text-based retrieval, recs-planations), not just item prediction. Severe loss of language retention degrades language steerability compliance and the quality of mixed text–SID outputs, even when ID-level retrieval remains competitive. By isolating SID integration to newly introduced token parameters before full instruction tuning, NEO achieves a better trade-off between item grounding and preserving general linguistic competence.

\paragraph{Is the framework backbone-agnostic?}
To demonstrate that the framework is backbone-agnostic, we did a focused adaptation of Llama 3.2 1B~\cite{grattafiori2024llama} with NEO and found that the Domain Grounding stage improves the performance by around 18\% in text-based retrieval, indicating that NEO is a framework-level contribution rather than a model-specific trick. All other results are observed in Llama as well. Since we are not focused on identifying the optimal backbone here, we limited the comprehensiveness of this evaluation; a comprehensive evaluation would also require careful selection of backbones that could be considered to be comparable by a number of other factors.

\subsubsection{Capability Induction. \label{sec:capability_induction}}
Finally, we verify that the model is capable of performing a diverse set of tasks from text-based retrieval, recommendation, text generation, the language steerability of the model, and discuss considerations for generalizing to new tasks.

\begin{table}[htb]
\small 
  \centering
  %\resizebox{\textwidth}{!}{ 
  \caption{Multi-task performance}
    \vspace*{-0.1in}
  \label{tab:main}
  \begin{tabularx}{\columnwidth}{X|ll|ll}
    \toprule[1.25pt]
    \multirow{2}{*}{\textbf{Method}} 
      & \multicolumn{2}{c|}{\textbf{HR@K$\uparrow$}} 
      & \multicolumn{2}{c|}{\textbf{NDCG@K$\uparrow$}} \\
    \cmidrule(lr){2-3}\cmidrule(lr){4-5}
      & \textbf{$k=10$} & \textbf{$k=30$} & \textbf{$k=10$} & \textbf{$k=30$} \\
    \midrule[1.1pt]
    \rowcolor[rgb]{0.93,0.93,0.93}
    \multicolumn{5}{c}{\texttt{Episode Recommendation}} \\
    \textcolor{gray}{\texttt{Baseline}}
      & \textcolor{gray}{-}
      & \textcolor{gray}{-}
      & \textcolor{gray}{-}
      & \textcolor{gray}{-}
    \\
    \texttt{NEO - mono}  
       & $\textcolor{scoregreen}{+57\%}$ 
       & $\textcolor{scoregreen}{+41\%}$ 
       & $\textcolor{scoregreen}{+80\%}$ 
       & $\textcolor{scoregreen}{+60\%}$
      \\
    \texttt{NEO - multi}  
      & $\textcolor{scoregreen}{+58\%}$ 
      & $\textcolor{scoregreen}{+40\%}$ 
      & $\textcolor{scoregreen}{+80\%}$ 
      & $\textcolor{scoregreen}{+59\%}$ 
      \\

    \midrule\rowcolor[rgb]{0.93,0.93,0.93}
    \multicolumn{5}{c}{\texttt{Show Recommendations}} \\
    \textcolor{gray}{Baseline} 
      & $\textcolor{gray}{-}$
      & $\textcolor{gray}{-}$
      & $\textcolor{gray}{-}$
      & $\textcolor{gray}{-}$
    \\
    
    \texttt{NEO - mono}  
      & $\textcolor{scoregreen}{+20\%}$ 
      & $\textcolor{scoregreen}{+2\%}$ 
      & $\textcolor{scoregreen}{+46\%}$ 
      & $\textcolor{scoregreen}{+30\%}$
      \\

    \texttt{NEO - multi} 
      & $\textcolor{scoregreen}{+24\%}$ 
      & $\textcolor{scoregreen}{+2\%}$ 
      & $\textcolor{scoregreen}{+58\%}$ 
      & $\textcolor{scoregreen}{+39\%}$ 
      \\

    \midrule
    \rowcolor[rgb]{0.93,0.93,0.93}
    \multicolumn{5}{c}{\texttt{Audiobook Recommendation}} \\
    \textcolor{gray}{Baseline} 
      & \textcolor{gray}{$-$ }
      & \textcolor{gray}{$-$ }
      & \textcolor{gray}{$-$ }
      & \textcolor{gray}{$-$ }
    \\

    \texttt{NEO - mono}
      & $\textcolor{scoregreen}{+36\%}$ 
       & $\textcolor{scoregreen}{+6\%}$ 
       & $\textcolor{scoregreen}{+73\%}$ 
       & $\textcolor{scoregreen}{+52\%}$
      \\
    
    \texttt{NEO - multi}  
      & $\textcolor{scoregreen}{+46\%}$ 
      & $\textcolor{scoregreen}{+14\%}$ 
      & $\textcolor{scoregreen}{+97\%}$ 
      & $\textcolor{scoregreen}{+66\%}$ 
      \\

    \midrule
    \rowcolor[rgb]{0.93,0.93,0.93}
    \multicolumn{5}{c}{\texttt{Text-based retrieval}} \\
    \textcolor{gray}{Baseline} 
    & \textcolor{gray}{0.45} & 
    \textcolor{gray}{0.51} 
    & \textcolor{gray}{0.37} 
    & \textcolor{gray}{0.38} 
    \\

    \texttt{NEO - mono} 
      & $\textcolor{scoregreen}{+45\%}$ 
      & $\textcolor{scoregreen}{+36\%}$ 
      & $\textcolor{scoregreen}{+62\%}$ 
      & $\textcolor{scoregreen}{+58\%}$ 
      \\
    \texttt{NEO - multi}  
      & $\textcolor{scoregreen}{+47\%}$ 
      & $\textcolor{scoregreen}{+36\%}$ 
      & $\textcolor{scoregreen}{+62\%}$ 
      & $\textcolor{scoregreen}{+58\%}$ 
      \\
    \bottomrule[1.25pt]
  \end{tabularx}%}
    \vspace*{-0.1in}
\end{table}

\paragraph{Multi-task Performance on Text-based Retrieval and Recommendation}
Table~\ref{tab:main} reports results for multiple recommendation as well as text-based retrieval performance cast as SIDs generation. Multi-task training yields small but consistent improvements over mono-task across HR@K and NDCG@K, indicating positive transfer from jointly training recommendation, retrieval, and explanation objectives. For text-based retrieval, performance varies by item type (see Table \ref{tab:search}), with episode retrieval being the most challenging due to finer granularity and higher lexical ambiguity. 
NEO is trained on only about 50\% of the historical logs used by the production baseline model, highlighting the sample-efficiency of our model.

\paragraph{Text Generation for Interest profiling and Recsplanations}
When asked to analyze a user’s listening history, represented as an ordered list of SIDs, the model can recover coherent high-level interests and produce item-grounded explanations. We quantify the quality of these generations using an LLM-as-a-judge protocol~\cite{fabbri2025evaluating, gu2024survey} with GPT-4o-mini. For interest profiling, the judge evaluates three dimensions: coverage, groundedness, and faithfulness. For recsplanations, we assess faithfulness to the recommended item, faithfulness to the user history, and deceptiveness (the tendency to introduce plausible but unsupported rationales). Each dimension is scored on a 0–5 scale. Full evaluation prompts are provided in \Cref{appendix:llmasajudge}. 

\Cref{tab:llmjudge} reports the results, showing that NEO enables the model to interpret SID-based user histories and leverage the semantic structure encoded by the identifier space to produce grounded natural-language outputs. We provide qualitative examples of interest profiles and recsplanations in \Cref{appendix:llmasajudge}.

\begin{table}[htb]
\small 
  \centering
  \caption{LLM as a Judge result for user analysis and recsplanation tasks. Scores from 0 to 5}
    \vspace*{-0.1in}
  \label{tab:llmjudge}
  %\resizebox{\textwidth}{!}{ 
  \begin{tabular}{@{}ll|cc@{}}
    \toprule[1.25pt]
    \multicolumn{2}{c|}{\textbf{Variation}}
      & \textbf{Artists} 
      & \textbf{Episodes}  \\
    \midrule[1.1pt]
    \multirow{2}{*}{Interest profiling} & Coverage 
    & 3.79 & 3.47
    \\
    & Groundedness  
    & 3.81 & 3.50
    \\
    \midrule
    \multirow{3}{*}{Recsplanation} & Faithfulness item & 4.08 & 3.28
    \\
    & Faithfulness history & 4.16 & 2.79
    \\
    & Deceptiveness & 4.74 & 4.52
      \\
    \bottomrule[1.25pt]
  \end{tabular}%}
    \vspace*{-0.1in}
\end{table}

\paragraph{Is the LLM item-type language-steerable?}
To evaluate language steering, we change only \texttt{Target Item Type: <item-type>} from each original data sample and disable the constrained decoding. 
The model successfully follows these instructions, steering all 30 beams toward the desired SIDs.
To improve few- and zero-shot generalization~\cite{brown2020language}, it will be critical for future work to expand the variety of tasks during this stage to increase capability induction, deferring task-specific behaviors to the final fine-tuning phase.

\section{Conclusions}

We introduced NEO, a tool-free framework for adapting decoder-only LLMs to operate over large, heterogeneous catalogs by interleaving natural language with semantic identifiers and enforcing catalog-valid generation via prefix-trie–constrained decoding. At a scale exceeding 10M items, NEO consistently outperforms strong production baselines for recommendation and text-based retrieval, while remaining language-steerable and supporting mixed ID–text outputs for explanations and user understanding.

Beyond empirical gains, our design analysis highlights two key principles for grounding LLMs in large discrete domains: the importance of high quality semantically structured identifiers over atomic IDs, and the role of a staged alignment recipe in balancing item grounding with retention of general language competence, key for tasks involving mixed text-SIDs in output. Our results show that naive alternatives—such as random tokenization, collapsing alignment stages, or continuous pre-training—either degrade performance or compromise language capabilities, especially for tasks that require instruction following and mixed text–item generation.

More broadly, we view NEO as an instance of a general approach to building fully self-contained, domain-adapted LLMs that reason jointly over entities, user behavior, and language without relying on external tools or task-specific pipelines. We believe the framework and empirical insights presented here provide practical guidance for extending LLMs to other large-scale discovery settings involving heterogeneous entities and steerable generative behavior.

\begin{acks}
Special thanks to Cara Van Uden, Matthew N.K. Smith, Paul Gigioli, David Murgatroyd, Vincent Tang, Doug J. Cole, Martin Bomio, Enrico Palumbo, Zahra Nazari, Yu Zhao, Francesco Fabbri, Ali Vardasbi, José Luis Redondo García, Gustavo Penha, Oguz Semerci, Sahitya Mantravadi, Alice Y. Wang, Catie Edwards, and Alyssa Yang. We also thank the many engineers, researchers, data scientists, and product managers across these groups whose feedback shaped the modeling, serving, and evaluation behind this work. 
\end{acks}

\bibliographystyle{acm}
\bibliography{biblio}

\balance
\appendix
\newpage

\section{Example Text Generation}\label{exampletextgeneration}
Figure \ref{fig:recsplanations} shows an example of a ``Recs-planation'' generated -- an explanation for the recommendation generated together with the SID. In this case, the portion indicated by ``Recommendation:'' is what is returned to the user with the semantic ID that was predicted resolved to a catalog item.

Figure \ref{fig:user_understanding} demonstrates how NEO can predict a user interest profile based on just a sequence of SIDs from the user's historical interactions. Evaluation of recs-planations and user profiling is reported in Section \ref{sec:capability_induction}

\begin{figure}[htb]
\captionsetup{skip=4pt}
\begin{resultbox}{Result: Recs-planations}
{\itshape
\red{USER: I'm looking for a funny podcast that explores moral theories. The user history is: \(\langle\)list of Semantic IDs\(\rangle\). Provide...}

\medskip
ASSISTANT: The user's podcast history suggests an interest in comedy and storytelling, with a focus on humorous conversations and absurd situations. They enjoy episodes with lighthearted and [\ldots] The user also appears to appreciate episodes that explore themes of friendship, adventure, and personal growth, particularly in the context of fantasy role-playing and comedy shows. 

\medskip
\inlinehl{Recommendation}: \(\langle\)Semantic ID\(\rangle\). This episode discusses morality in a humorous way, aligning with your interest in comedy and thought-provoking discussions.}

\end{resultbox}
\caption{Recsplanation result for an episode request. This randomly selected result shows that the model can interleave recommendations (SIDs) and explanation (text) fluently.}
\label{fig:recsplanations}
\end{figure}

\begin{figure}[htb]
\captionsetup{skip=4pt}
\begin{resultbox}{Result: user understanding}
{\itshape
\red{USER: Build a listener profile: What kind of podcasts does this user enjoy? What themes or characteristics define their taste? The user history is: \(\langle\)list of Semantic IDs\(\rangle\). Provide...}

\inlinehl{Ground truth:} Pronounced interest in women\'s health, particularly hormonal wellness, fertility, and healthcare disparities, as seen in their choice of a dedicated episode on women\'s health topics. They also consistently engage with holistic wellness practices and personal development content centered on entrepreneurship, mindset, and women\'s leadership. [\ldots]
}

\medskip
\inlinehl{Prediction from SIDs}: Clear preference for content centered on holistic health and women's empowerment. Their listening history consistently explores resilience through health (PCOS, PCOS treatment), mental well-being (coping with loss, self-worth), and practical self-improvement strategies. They engage deeply with narratives of overcoming health challenges, redefining success, and integrating physical and emotional wellness into daily life. [\ldots]
\end{resultbox}

\caption{Qualitative example of user understanding for an episode, from the SID history of an user.}
\label{fig:user_understanding}
\end{figure}

\section{Implementation details}

\begin{table}[h]
\centering
\small
\caption{Detailed hyperparameters of NEO applied to Qwen}
\label{tab:hyper}
\begin{tabular}{l|cc}
\toprule
\textbf{Hyperparameters} & \textbf{Stage 2} & \textbf{Stage 3}  \\
\midrule
Learning rate & $1.0\times10^{-3}$   & $1\times10^{-4}$  \\
LR scheduler  & Cosine & Cosine  \\
Weight decay  & $0.01$ & $0.01$  \\
Gradient clip & $1.0$ & $1.0$  \\
Optimizer     & \multicolumn{2}{c}{AdamW ($\beta_1=0.9, \beta_2=0.95$)} \\
Warm-up steps    & $300$      & $400$  \\
Batch size       & $64$      & $64$   \\
\bottomrule
\end{tabular}
\end{table}

For an efficient training, we use sequence packing and Ray, which helps distributing the load between CPUs and GPUs. We use PyTorch and 8x Nvidia H100 GPUs. \Cref{tab:hyper} shows the hyperparameters used for training

\subsection{Evaluation metrics}
\label{sec:metrics}
We evaluate the performance of our recommendation task on implicit feedback through two standard metrics, namely HR@K and NDCG@K.

\begin{equation}
\text{HR}@K = \frac{\sum_{u \in U}{\mathds{1}(\text{the relevant item is in top } K)}}{|U|}
\label{eq:HR}
\end{equation}

\begin{equation}
\text{NDCG}@K = \frac{1}{|U|} \sum_{u \in U} \frac{\log(2)}{\log(r_u + 1)}
\label{eq:NDCG}
\end{equation}

where $U$ is the set of users, $r_u$ is the rank of the relevant item for user $u$ (if the item is within the top $K$, otherwise the contribution is 0), and $\Upsilon$ is the entire catalogue.

\subsection{Podcast recommendations}
Table~\ref{tab:episode_recommendations} reports podcast episode recommendation performance broken down by user type into coldstart and warmstart. Performance gains are larger for warm-start users (i.e., users with non-empty interaction histories), where richer histories provide stronger personalization signals. However, there is room for improvement for simulated cold-start users, where the baseline model uses specialized side features specifically designed for the no-history regime.

\begin{table}[htb]
  \centering
  %\resizebox{\textwidth}{!}{ 
  \begin{tabular}{l|cc|cc}
    \toprule[1.25pt]
    \multirow{2}{*}{\textbf{User type}} 
      & \multicolumn{2}{c|}{\textbf{HR@K$\uparrow$}} 
      & \multicolumn{2}{c|}{\textbf{NDCG@K$\uparrow$}} \\
    \cmidrule(lr){2-3}\cmidrule(lr){4-5}
      & \textbf{$k=10$} & \textbf{$k=30$} & \textbf{$k=10$} & \textbf{$k=30$} \\
    \midrule[1.1pt]
    Coldstart 
    & $\textcolor{scorered}{-48\%}$ 
    & $\textcolor{scorered}{-57\%}$ 
    & $\textcolor{scorered}{-34\%}$ 
    & $\textcolor{scorered}{-40\%}$ 
    \\
    Warmstart 
    & $\textcolor{scoregreen}{+57\%}$ 
    & $\textcolor{scoregreen}{+40\%}$ 
    & $\textcolor{scoregreen}{+70\%}$ 
    & $\textcolor{scoregreen}{+60\%}$ 
    \\
    \bottomrule[1.25pt]
  \end{tabular}%}
  \caption{Episode recommendations detailed results. The LLM has room for improvement in coldstart users, which represent $\sim$ 2\% of our data.}
  \label{tab:episode_recommendations}
\end{table}

\subsection{Search by item type and query complexity}
\Cref{tab:search} further analyzes a subset of challenging queries that required user reformulation ("Multi-steps"); these sessions are difficult to construct because the initial query did not immediately lead to engagement and remain an open challenge for single-shot retrieval.

Our results show that the significantly outperform the baseline model, especially in multi-steps settings.

\begin{table}[htb]
  \centering
  \caption{Text-based retrieval/search performance by item type and query complexity.}
  \label{tab:search}
  %\resizebox{\textwidth}{!}{ 
  \begin{tabular}{l|ll|ll}
    \toprule[1.25pt]
    \multirow{2}{*}{\textbf{Variation}} 
      & \multicolumn{2}{c|}{\textbf{HR@K$\uparrow$}} 
      & \multicolumn{2}{c|}{\textbf{NDCG@K$\uparrow$}} \\
    \cmidrule(lr){2-3}\cmidrule(lr){4-5}
      & \textbf{$k=10$} & \textbf{$k=30$} & \textbf{$k=10$} & \textbf{$k=30$} \\
    \midrule[1.1pt]
    \rowcolor[rgb]{0.93,0.93,0.93}
    \multicolumn{5}{c}{\texttt{Item type}} \\
    Artists 
    & $\textcolor{scoregreen}{+36\%}$ 
    & $\textcolor{scoregreen}{+35\%}$ 
    & $\textcolor{scoregreen}{+38\%}$ 
    & $\textcolor{scoregreen}{+38\%}$ 
    \\
    Audiobooks  
    & $\textcolor{scoregreen}{+26\%}$ 
    & $\textcolor{scoregreen}{+27\%}$ 
    & $\textcolor{scoregreen}{+29\%}$ 
    & $\textcolor{scoregreen}{+30\%}$ 
    \\
    Episodes 
    & $\textcolor{scoregreen}{+40\%}$ 
    & $\textcolor{scoregreen}{+16\%}$ 
    & $\textcolor{scoregreen}{+73\%}$ 
    & $\textcolor{scoregreen}{+59\%}$ 
    \\
    \rowcolor[rgb]{0.93,0.93,0.93}
    \multicolumn{5}{c}{\texttt{Query complexity}} \\
    One step
    & $\textcolor{scoregreen}{+21\%}$ 
    & $\textcolor{scoregreen}{+12\%}$ 
    & $\textcolor{scoregreen}{+33\%}$ 
    & $\textcolor{scoregreen}{+30\%}$ 
    \\
    Multi-steps
    & $\textcolor{scoregreen}{+185\%}$ 
    & $\textcolor{scoregreen}{+176\%}$ 
    & $\textcolor{scoregreen}{+243\%}$ 
    & $\textcolor{scoregreen}{+229\%}$ 
      \\
    \bottomrule[1.25pt]
  \end{tabular}%}
\end{table}

\section{Semantic ID Formal Details \label{appendix:semid_formal}}
Formally, given an item embedding $x_e \in \mathbb{R}^d$, we map it to an SID defined as
\begin{equation}
\mathrm{SID}(e) = (c_1, c_2, \ldots, c_M), \qquad c_m \in \{1,\ldots,K\}.
\label{eq:sid_def}
\end{equation}
where $K$ denotes the number of clusters per position $M \ll K$ to ensure meaningful semantic partitioning. Thus, each SID is an ordered tuple of discrete codes drawn from $M$ codebooks.

\paragraph{Residual Quantization} NEO is agnostic to a specific quantizer method used to construct SIDs. However, in this work we focus on residual quantizers, which often induces a coarse-to-fine decomposition: earlier codes capture broad semantic regions, and later codes refine within-region distinctions. This property aligns naturally with autoregressive generation, where tokens are produced sequentially.
Specifically, we adopt Residual $K$-means~\cite{ju2025generative} for its flexibility and empirical stability, particularly compared to alternatives such as RQ-VAE, which often require multiple auxiliary losses to avoid poor local minima~\cite{lancucki2020robust, mentzer2023finite, huh2023straightening}. 

Residual $k$-means represents each embedding as a sequence of codebook indices by iteratively quantizing the residual error. Concretely, we maintain $M$ codebooks $\{C_1,\ldots,C_M\}$, each containing $K$ centroids $\{\mu_{m,1},\ldots,\mu_{m,K}\}$. Starting from the initial residual $r_0 = x_e$, we select the closest centroid at each stage
\begin{align}
c_m &= \arg\min_{j \in \{1,\ldots,K\}} \|r_{m-1} - \mu_{m,j}\|_2^2, \label{eq:rkmeans_assign}
\end{align}
and update the residual as:
\begin{align}
r_m &= r_{m-1} - \mu_{m,c_m}, \qquad m=1,\ldots,M. \label{eq:rkmeans_residual}
\end{align}

\paragraph{Choosing SID Length and Codebook Size} The SID length and codebook size $K$ trade off addressability, semantic coherence, and sequence length. Longer SIDs increase addressability but inflate prompt length, while shorter SIDs risk semantic collisions.  We guide this choice using two criteria: (i) collision rate, the fraction of SID tuples assigned to multiple items; and (ii) prefix coherence, the average cosine similarity between items sharing the same SID prefix. We select $M$ and $K$ to minimize collisions while maintaining high semantic coherence for short prefixes, enabling reliable generation without excessive sequence overhead.

\paragraph{Item embeddings from content}
\label{sec:embeddings}
The quality of SIDs depends critically on the underlying item embedding $x_e$. Such 
embeddings may be derived from collaborative signals~\cite{wang2024content, wang2024learnable}, content metadata~\cite{rajput2024recommender, singh2024better}, or hybrid representations~\cite{2025_PLUMAdaptingPretrainedLanguageModels, kim2024sc, 2025_EAGERLLMEnhancingLargeLanguageModels}, provided they induce a meaningful semantic geometry over items. 

In practice, item representations must remain stable under distribution shift. In domains with strong seasonality and rapidly evolving trends (e.g., music, video, or spoken audio), embeddings derived from collaborative signals can drift quickly, causing an item to be reassigned different identifiers over short time spans. This drift may require frequent retraining of both LLM and the quantizer, or the introduction of additional stabilization mechanisms. Moreover, collaborative signals are often sparse or unavailable for cold-start items. 
For these reasons, NEO constructs SIDs using content-based item representations, and defers delegates the incorporation of collaborative signals to later instruction-tuning stages, where the LLM can learn to leverage user interaction history directly.

\section{CF-based Semantic IDs stability}
\label{appendix:jaccard}
We have analyzed the stability of our CF-based embeddings by sampling 1{,}000{,}000 random artists, computed each artist's top-50 nearest neighbors under the CF embedding space, and measured neighborhood overlap across time using Jaccard similarity.
Over a 3-week horizon, the median overlap is 0.59 (Jaccard@50), corresponding to roughly 37 out of 50 neighbors retained; the least stable decile has overlap 0.37, corresponding to roughly 27 out of 50 neighbors retained.

CF-based SIDs may exhibit temporal instability due to evolving user behavior. This instability suggests that CF-based tokenizations may require frequent retraining of the quantizer and the LLM, or explicit stabilization mechanisms (e.g., anchoring or temporal regularization), to function as reliable identifiers in fast-moving catalogs.

\begin{figure}
\captionsetup{skip=4pt}
\begin{InstructionBox}[label={prompt:user_analysis}]{LLM as a Judge (user)}
You are an expert evaluator of user-interest summaries derived from podcast listening history.\\
Your job: judge how well the prediction matches the ground\_truth summary of user interests, USING the user\_history\_with\_descriptions as evidence.\\
\\
Be strict about grounding:\\
- Do NOT reward eloquence; reward factual and thematic alignment.\\
- Do NOT assume extra user info beyond what is in the user\_history\_with\_descriptions.\\
\\
Evaluate across these dimensions (0–5 each):\\
1) COVERAGE: Does the prediction capture the main themes from ground\_truth, or user\_history\_with\_descriptions)?\\
2) EVIDENCE GROUNDEDNESS: Are prediction claims supported either by either the user\_history\_with\_descriptions or ground\_truth?\\
\\
Then produce:\\
- average\_score (0–5)\\
- verdict: a small textual verdict\\
Return ONLY valid JSON with this schema:\\
\\
    "coverage": int,\\
    "evidence\_groundedness": int,\\
    "oaverage\_score": int\\
  "verdict": string,\\

\end{InstructionBox}
\caption{Prompt template used for LLM as a Judge for the user analysis task.}
\label{fig:llmjudge_user}
\end{figure}

\begin{figure}
\captionsetup{skip=4pt}
\begin{InstructionBox}[label={prompt:recsplantions}]{LLM as a Judge (user)}
You are an expert evaluator of EXPLANATIONS produced by a podcast recommendation system.\\
\\
You will be given:\\
- user\_history\_with\_descriptions: a list of recently listened episode descriptions (text)\\
- suggested\_episode: the recommended episode title/description (text)\\
- explanation: the system's natural-language explanation for why it recommended the episode. It is based to align the user to stream the item\\
\\
Your task: judge whether the explanation is (a) faithful to the suggested\_episode, and (b) useful to the user.\\
\\
CRITICAL RULES:\\
- Use ONLY user\_history\_with\_descriptions and suggested\_episode as evidence.\\
- Do NOT assume hidden signals (e.g., demographics, embeddings, similar users, popularity, “because you liked X” unless X is in user\_history\_with\_descriptions).\\
- The explanation must connect at least one concrete aspect of user\_history\_with\_descriptions to one concrete aspect of the suggested\_episode.\\
\\
Score each dimension 0–5:\\
1) FAITHFULNESS\_TO\_HISTORY: Are references to user interests grounded in user\_history\_with\_descriptions?\\
2) FAITHFULNESS\_TO\_ITEM: Are claims about the recommended episode grounded in suggested\_episode text?\\
3) NON\_DECEPTIVE: Avoids implying the system used signals not present (“people like you”, “trending”, “based on your location”) unless evidenced.\\
\\
Compute:\\
overall\_score (0–5)\\
\\
Then produce:
- verdict (1–2 sentences)\\
\\
Return ONLY valid JSON:\\
    "faithfulness\_to\_history": int,\\
    "faithfulness\_to\_item": int,\\
    "non\_deceptive": int,\\
    "overall\_score": int\\
  "verdict": string,\\

\end{InstructionBox}
\caption{Prompt template used for LLM as a Judge for the recsplanation task.}
\label{fig:llmjudge_recsplanation}
\end{figure}

\section{LLM as a Judge}
\label{appendix:llmasajudge}
\Cref{fig:llmjudge_user,fig:llmjudge_recsplanation} shows the prompts used to evaluate our recsplanations and interest profiling from SID sequences. \Cref{fig:recsplanations,fig:user_understanding} instead shows some qualitative results.

\section{Unsuccessful attempts}
\textbf{SID expert.} Following BEIT-3~\cite{wang2023image}, we modified the LLM backbone architecture to add a SID expert, activated with a hard router mask whenever the input token was a SID. The performance was similar to the non-expert approach, so we have discarded this modification.

\textbf{Vision as LoRA.} We implemented Vision as Lora~\cite{wang2025vision}, which proves to be effective in extending (via LoRA~\cite{hu2022lora} text-only LLMs for the new vision modality. However, 
we have observed lower performance than NEO.

\textbf{Semantic ID token embeddings for k-NN.} 
To connect NEO-trained generative models with embedding-based retrieval, we explored whether the learned SID token embeddings can be used to construct an item embedding space for approximate $k$-NN. After training Qwen3-0.6B with NEO, we represent each item by the concatenation (or mean) of the $M$ learned input embeddings corresponding to its SID tokens. We then build a simple episode-retrieval recommender where each user is represented by the average embedding of episodes streamed in the previous 30 days, and recommendations are obtained via approximate $k$-NN in the resulting embedding space.

In our experiments, this approach significantly underperforms the production retrieval baseline and the generative NEO model. A likely explanation is that concatenating token embeddings yields very high-dimensional item vectors (e.g., $M{=}4$ and hidden size 1024 $\Rightarrow$ 4096 dimensions), which can make nearest-neighbor retrieval less reliable without additional metric learning or dimensionality reduction. We also evaluated alternative text-based embedding models (e.g., WordLlama~\cite{miller2024wordllama}) for item representations, but did not observe improvements. We leave a more principled study of deriving retrieval-friendly embeddings from SID-token representations (e.g., projection layers, pooling/whitening, contrastive fine-tuning, or mixed models~\cite{llm2vec}) to future work.

\section{Prompt templates}
\label{appendix:prompts}
%\Cref{fig:prompt_audiobook}, \Cref{fig:prompt_episodes}, \Cref{fig:prompt_shows} and \Cref{fig:prompt_search} show the prompt templates used for our experimental tasks.
\Cref{fig:prompt_search}, \Cref{fig:prompt_episodes}, \Cref{fig:prompt_shows} and \Cref{fig:prompt_search} show the prompt templates used for our experimental tasks.

\begin{figure}[t]
\captionsetup{skip=4pt}
\begin{InstructionBox}[label={prompt:search}]{Text-based Search/retrieval}
{\itshape
\inlinehl{USER:} 
You are a specialized Search Assistant for music and talk audio.
Your goal is to retrieve the most relevant artist, podcast episode, or audiobook based on the user's query and listening history.

\medskip
*** Instructions\\
1. Analyze the user's search query and country.\\
2. Use the user's affinity history (most listened) for personalization context.\\
3. Return the specific Semantic ID for the requested entity type.

\medskip          
*** User Context\\
<SIDs of the most liked artists, episodes and audiobooks, if present>

\medskip
*** Current Request\\
User Country (ISO 3166-1 alpha-2 code): \(\langle\)user\_country\(\rangle\)\\
Search Query: \(\langle\)query\(\rangle\)

\medskip
*** TARGET\\
Based on the query and context, output the Semantic ID for the following item type only.\\
Target Entity Type: **<item\_type>**
}

\medskip
\inlinehl{ASSISTANT}: \(\langle\)Semantic ID\(\rangle\)[the semantic id]\(\langle\)/Semantic ID\(\rangle\)

\end{InstructionBox}
\caption{Prompt template used for text-based retrieval/search.}
\label{fig:prompt_search}
\end{figure}

\begin{figure}[b]
\captionsetup{skip=4pt}
\begin{InstructionBox}[label={prompt:familiarepisodes}]{Episodes}
{\itshape
\inlinehl{USER:} You are an expert in podcast recommendations. Your task is to suggest a new episode the user would like to stream from the user's podcast listening habits based on the following information: user history, registration country, language preferences. Use the information you need to make the best recommendation. The user history might be empty if the user is a new user.

\medskip
**Constraint** \\
- Start the answer with the semantic ID, thus with \(\langle\)Semantic ID\(\rangle\) and end it with \(\langle\)/Semantic ID\(\rangle\) 

\medskip
**User's information** \\
\(\langle\)list of streamed SIDs\(\rangle\) \\
The user is registered in (country code): \(\langle\)registration country\(\rangle\)  \\
The user listens to content in the following languages: \(\langle\)languages\(\rangle\) \\
Recommend the next item the user would like to stream.
}

\medskip
\inlinehl{ASSISTANT}: \(\langle\)Semantic ID\(\rangle\)\(\langle\)the semantic id\(\rangle\) \(\langle\)/Semantic ID\(\rangle\)

\end{InstructionBox}
\caption{Prompt template used for episode recommendation.}
\label{fig:prompt_episodes}
\end{figure}

%\clearpage
\begin{figure}[h]
\captionsetup{skip=4pt}
\begin{InstructionBox}[label={prompt:shows}]{Shows}
{\itshape
\inlinehl{USER:} As a podcast specialist, predict which show the user will stream next.\\
Analyze the user's profile below to identify patterns in their listening behavior. \\
Note that new users may have limited or no history available.\\
\medskip
**User Profile**\\
\(\langle\)artist followed SIDs\(\rangle\) \\
\(\langle\)audiobook streams SIDs\(\rangle\) \\
\(\langle\)shows user history SIDs\(\rangle\) \\
\(\langle\)recent show streams SIDs\(\rangle\) \\
Country (ISO code): \(\langle\)registration country\(\rangle\) \\
Preferred languages (ISO codes): \(\langle\)languages\(\rangle\) \\
}
\medskip
Based on this profile, which show should be recommended?\\

\medskip
\inlinehl{ASSISTANT}: \(\langle\)Semantic ID\(\rangle\)[the semantic id]\(\langle\)/Semantic ID\(\rangle\)

\end{InstructionBox}
\caption{Prompt template used for show recommendation.}
\label{fig:prompt_shows}
\end{figure}

\begin{figure}[h]
\captionsetup{skip=4pt}
\begin{InstructionBox}[label={prompt:audiobooks}]{Audiobooks}
{\itshape
\inlinehl{USER:} You are an expert in audiobook recommendations. Your task is to suggest a new audiobook the user would like to stream based on the user's listening habits and preferences. Consider the following information: user history, registration country, and language preferences. Use the information you need to make the best recommendation. The user history might be empty if the user is a new user.\\
\medskip
**Constraint**\\
- Start the answer with the semantic ID, thus with \(\langle\)Semantic ID\(\rangle\) \\
and end it with \(\langle\)/Semantic ID\(\rangle\)\\
\medskip
**User's information**\\
\(\langle\)description of the contextual content (shows) the user has listened to\(\rangle\) \\
Here are the user's last \(\langle\)number\(\rangle\)  historical interactions, ordered by timestamp: \(\langle\)Semantic ID\(\rangle\)\(\langle\)list of SIDs\(\rangle\)\(\langle\)/Semantic ID\(\rangle\)\\
The user is registered in (country code): \(\langle\)registration country\(\rangle\) .\\
The user listens to content in the following languages (ISO codes): \(\langle\)languages\(\rangle\) .\\
The user has listened to audiobooks in the following genres: \(\langle\)genres\(\rangle\) \\
Recommend the next audiobook the user would like to stream.\\
}

\medskip
\inlinehl{ASSISTANT}: \(\langle\)Semantic ID\(\rangle\)[the semantic id]\(\langle\)/Semantic ID\(\rangle\)

\end{InstructionBox}
\caption{Prompt template used for audiobook recommendation.}
\label{fig:prompt_audiobook}
\end{figure}

\section{LSH}
\label{appendix:LSH}
For any embedding vector $\mathbf{x} \in \mathbb{R}^{d_{embed}}$, a hash function $h_{r}(\mathbf{x})$ is defined as:
\begin{equation}
h_{r}(\mathbf{x}) = \text{sgn}(\mathbf{w} \cdot \mathbf{x})
\end{equation}
where $\mathbf{w} \in \mathbb{R}^{d_{embed}}$ is a random hyperplane vector sampled from a standard Gaussian distribution $\mathcal{N}(0, I)$. To improve precision and reduce false positives, we concatenate $K$ such independent hash functions to form a $K$-bit signature:
\begin{equation}
H(\mathbf{x}) = [h_{1}(\mathbf{x}), h_{2}(\mathbf{x}), \dots, h_{K}(\mathbf{x})]
\end{equation}
The probability that two vectors $\mathbf{u}$ and $\mathbf{v}$ result in the same hash bit for a single random projection is a function of the angle $\theta$ between them:
\begin{equation}
P(h_{r}(\mathbf{u}) = h_{r}(\mathbf{v})) = 1 - \frac{\theta(\mathbf{u}, \mathbf{v})}{\pi}
\end{equation}

\end{document}